\documentclass[iop,apjl]{emulateapj}
\usepackage{graphicx}
\usepackage{amssymb}
\usepackage{natbib}
\usepackage{color}
\bibpunct{(}{)}{;}{a}{}{,}

\begin{document}

\newcommand{\etal}{{\rm ~et al.}}
\newcommand{\kms}{\ifmmode{{\rm ~km~s}^{-1}}\else{~km~s$^{-1}$}\fi}
\newcommand{\ms}{M$_{\odot}$}
\newcommand{\ls}{L$_{\odot}$}
\newcommand{\msyr}{M$_{\odot}$~yr$^{-1}$}
\newcommand{\pas}{$\rlap{.}^{\prime\prime}$}
\newcommand{\amm}{NH$_3$}

\font\cmss=cmss10 scaled 1200
\font\cmssbx=cmssbx10 scaled 1200

 \def\red#1 {\textcolor{red}{#1}~}
 \def\cyan#1 {\textcolor{cyan}{#1}~}

\shortauthors{Greenhill et al.}
\shorttitle{Magnetocentrifugal Wind}

\title{Dynamical Evidence for a Magnetocentrifugal Wind from a 20$\,M_\odot$ Binary Young Stellar Object
}

\author{
L. J. Greenhill,\altaffilmark{1}
C. Goddi,\altaffilmark{2}
C. J. Chandler,\altaffilmark{3}
L. D. Matthews,\altaffilmark{4}
and
E. M. L. Humphreys\altaffilmark{2}
}

\altaffiltext{1}{Harvard-Smithsonian Center for Astrophysics, 60 Garden 
Street,
Cambridge, MA 02138 USA; greenhill@cfa.harvard.edu.}

\altaffiltext{2}{European Southern Observatory, 
Karl-Schwarzschild-Strasse 2
D-85748 Garching bei M\"unchen, Germany}

\altaffiltext{3}{NRAO, P.O. Box O, Socorro, NM 87801}

\altaffiltext{4}{MIT Haystack Observatory, Off Route
   40, Westford, MA 01886}
\begin{abstract}

In Orion BN/KL, proper
motions of $\lambda 7$\,mm vibrationally-excited SiO masers trace  
rotation of a nearly edge-on disk and a bipolar wide-angle outflow 10-100\,AU 
from radio Source~I, a binary young stellar object (YSO) of $\sim$20\,M$_\odot$.   
Here we  map ground-state $\lambda7$\,mm SiO emission with the 
Very Large Array and track proper motions over 9 years. The innermost and strongest emission lies in
two extended arcs bracketing Source~I.   The proper motions trace
a northeast-southwest bipolar outflow 100-1000\,AU from Source~I with a median 
3D motion of  $\sim$18\,\kms.  An overlying distribution of $\lambda1.3$\,cm 
H$_2$O masers betrays similar flow characteristics.   Gas dynamics and 
emission morphology traced by the masers suggest the presence of 
a magnetocentrifugal disk-wind.  Reinforcing evidence lies in the colinearity 
of the flow, apparent rotation across the flow parallel to the  disk rotation, and recollimation  
that narrows the flow opening angle $\sim  120$\,AU downstream.  The
arcs of ground-state SiO emission may mark the transition point to a shocked super-Alfv\'enic outflow.

\end{abstract}


\section{Introduction}

The balance of gravitational, radiative, and magnetic forces driving high-mass star formation is poorly understood, in part because it has not been possible in general to resolve regions where outflows are launched and collimated. Complicating study, high-mass young stellar objects (YSOs)  are deeply embedded during the accretion phase, evolve rapidly, and tend to form in distant crowded regions for which observations may be confusion-limited.

The nearest high-mass YSO, radio Source~I in  Orion BN/KL ($418\pm6$\,pc; Kim et al. 2008\nocite{kim08}) offers 
unique opportunities for investigation.  It is deeply embedded \citep{g04} in a crowded region \citep{gezari98, shuping04}.  However, it is surrounded by a compact ionized disk with R$\sim$40 AU resolved in the radio continuum (Goddi et al. 2011a), interpreted as either a hypercompact-HII region at T$\sim$8000 K emitting $p/e$ Bremsstrahlung around a $\sim$10 \ms\ YSO or a massive disk at $T< 4500$ K emitting via H$^-$ opacity and heated by $\sim10^5$ \ls\ \citep{reid07,plambeck2013}. \citet{goddi11a} have estimated a robust dynamical mass of  $\sim20$\,M$_\odot$ in an  equal-mass binary, favoring $p/e$ Bremsstrahlung.

Gas dynamical study is enabled by  an unusually large number of maser transitions of SiO and H$_2$O excited by the YSO \citep[e.g.,][]{goddi09,g98}.   Specifically, the position-velocity structure of vibrationally-excited SiO masers at projected radii of 10--100\,AU, resolved with very long baseline interferometry, outlines the limbs of a nearly edge-on, $\sim$14\,AU thick obscuring disk and a bipolar wide-angle outflow oriented northeast-southwest \citep{Gre04b, kim08, matthews10}.    Maser proper motions  clearly trace rotation and expansion in a disk/outflow \citep{matthews10}.  

Here, we analyze angular distributions and time evolution  for ground-state $\lambda 7$\,mm SiO and $\lambda 1$\,cm 
H$_2$O maser emission around Source~I.  The masers sample outflow on scales up to 1000\,AU, reinforcing the disk-outflow model, and provide among the best dynamical evidence thus far of a magnetocentrifugal disk-wind \citep{bp82,kp00} associated with a high-mass YSO.  

\section{Observations}

We observed SiO and H$_2$O maser emission toward Source~I with the
Very Large Array (VLA) of the National Radio Astronomy
Observatory\footnote{The NRAO is a facility
   of the National Science Foundation operated under cooperative
   agreement by Associated Universities, Inc.} at multiple epochs over
9 years (Table\,\ref{tab:obslog}).

{\bf SiO--} We correlated two simultaneous, single-polarization
basebands per epoch, one tuned to the $v$=0 transition ($\nu_{\rm
   rest}=43423.79$\,MHz) and the other to the much stronger $v$=1 
transition
($\nu_{\rm rest}=43122.08$\,MHz).  
3C286 or 3C48 were used as absolute flux calibrators; 0530+135 or 3C84 were
used as bandpass calibrators.  A 6.25\,MHz bandwidth covered
V$_{\rm lsr}$= -13.7 to 29.4\kms\ toward Source~I, with 97.656\,kHz (0.65\,km\,s$^{-1}$)
channel spacing.

We selected a strong $v$=1 Doppler component as a reference to
self-calibrate antenna gain and tropospheric fluctuations on 10$^s$
time-scales.  Scans of J0541--056 enabled calibration of slowly-varying 
phase offsets between the signal paths for the two observing
bands every 15-30$^m$, which enabled us to transfer the antenna and
tropospheric calibration to the band containing the (weaker) $v$=0
line (see Goddi\etal~2009\nocite{goddi09}).  

We imaged a region within $\pm 5''$ of Source~I.  Because $v$=0
emission contains both  extended and compact maser components ($T_b^{peak}\sim4\times10^6$\,K),
we used uniform ({\it u,v}) weighting to isolate compact knots and estimate
proper motions.  For other purposes, we used ROBUST=0 weighting
in AIPS (Table\,\ref{tab:obslog}).

\begin{deluxetable}{lcc@{\extracolsep{-0.05in}}ccc}
\tablecaption{Summary of Observations }
\tablehead{
\colhead{Date} &
\colhead{Project} &
\colhead{Array\tablenotemark{(a)}}   &
\colhead{Beam} &
\colhead{RMS}  \\
\colhead{(yymmdd)} &
\colhead{} &
\colhead{} &
\colhead{(mas) ~~ ($^\circ$)~ } &
\colhead{(mJy)} 
}

\startdata
\multicolumn{3}{l}{\hspace{-0.07in}$^{28}$SiO ${\rm J}=1\rightarrow0$ 
v=0, 1}  & \\
1999.08.28 & AG\,575 & A+~~~ &  $ 61 \times 43 ~@ +44^\circ$ & 10-40 \\
2002.03.31 & AG\,622 & A+~~~ &  $ 55 \times 45 ~@ -31^\circ$  & 6-20  \\
2006.04.15 & AC\,817 & A+~~~ &  $ 53 \times 39 ~@ -3.8^\circ$ & 2.5-8.5  \\
2009.01.12 & AG\,815 & A~~~~~ & $ 53 \times 43 ~@ +3.6^\circ$ & 3-20  \\
\multicolumn{3}{l}{\hspace{-0.07in}H$_2$O $6_{16}-5_{23}$} \\
2001.01.23 & AG\,578 & A+~~~ & $  113 \times 48 ~@ +30^\circ$ & $>6$  \\
\enddata

\tablenotetext{(a)}{A: 35 km maximum baseline; A+: 73 km maximum 
baseline via addition of the Pie-Town antenna.}
 
\label{tab:obslog}
\end{deluxetable}

\begin{figure}[tb]
\plotone{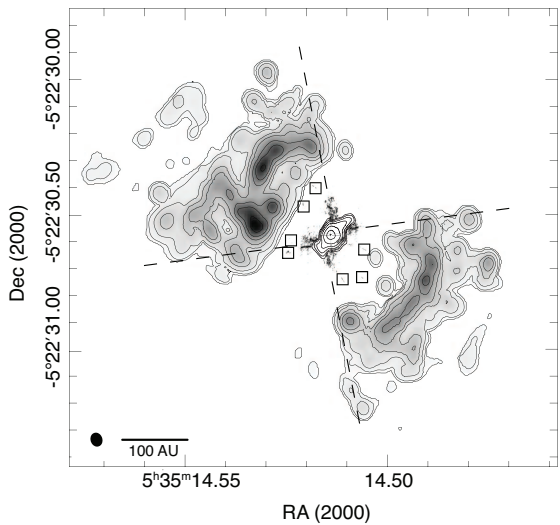}
\caption{Nested tracers of gas surrounding Source~I as observed at $\lambda$7~mm.
Velocity-integrated emission from the $v$=0 $J=1\rightarrow$0 transition of SiO (this work; epoch 2002.25) is shown in grayscale (26\,Jy bm$^{-1}$\,km\,s$^{-1}$ peak) with overlaid logarithmic contours ($2^N$).  Compact contours at the center depict $\lambda$7~mm continuum emission \citep{goddi11a}.  Black ``fuzz'' extending beyond these contours depicts the distribution SiO $v$=1,2 maser emission as imaged with the VLBA \citep{matthews10}.   The bulk of this emission lies in densely-sampled loci (``arms'') within $\sim$50~AU of Source~I, but isolated clumps (outlined in boxes for clarity) are found out to $\sim$80~AU. Dashed lines extend the four SiO $v$=1,2 arms to highlight that the $v$=0 emission subtends the same opening angle as structures on smaller scales.}

\label{fig:sio_integ_intensity}
\end{figure}

We tracked proper motions for 457 maser spots for between 2 and 4 epochs.  
To estimate proper motions, we  searched for
maser spots stronger than $5\sigma$ within each channel-map and fit each with a
two-dimensional elliptical Gaussian to obtain  position, flux-density, and angular size.  
Images of $v$=0 emission are
noise-limited, and relative position errors are given by
$0.5\frac{\theta}{SNR}$, where $\theta$ is beamwidth and SNR is the
peak intensity divided by the RMS noise in each velocity channel. 
Uncertainties for moderately bright emission were a few mas.
Cross-referencing of maser spots among different epochs could be done
by eye because the structure of the emission in each channel 
persisted with shifts of $<1$ beamwidth.  Proper motions were  calculated  using an
error-weighted linear least-squares fit to the fitted positions.   To
correct for motion of the reference  $v$=1 component,  we
computed proper motions relative to the strong $v$=0 feature at
+2.7\,\kms  
and then subtracted the mean motion of all those measured ($6.11\pm0.02$\kms~in right-ascension, $23.26\pm0.04$\kms~in declination).

We inferred absolute astrometry by measuring separation from BN,  in frequency-averaged images.  This agreed with that obtained using fast-switching to J0541-0541.  The estimated absolute position uncertainty is $\sim$3\,mas, (based on this comparison).

\begin{figure*}[!h]
\plotone{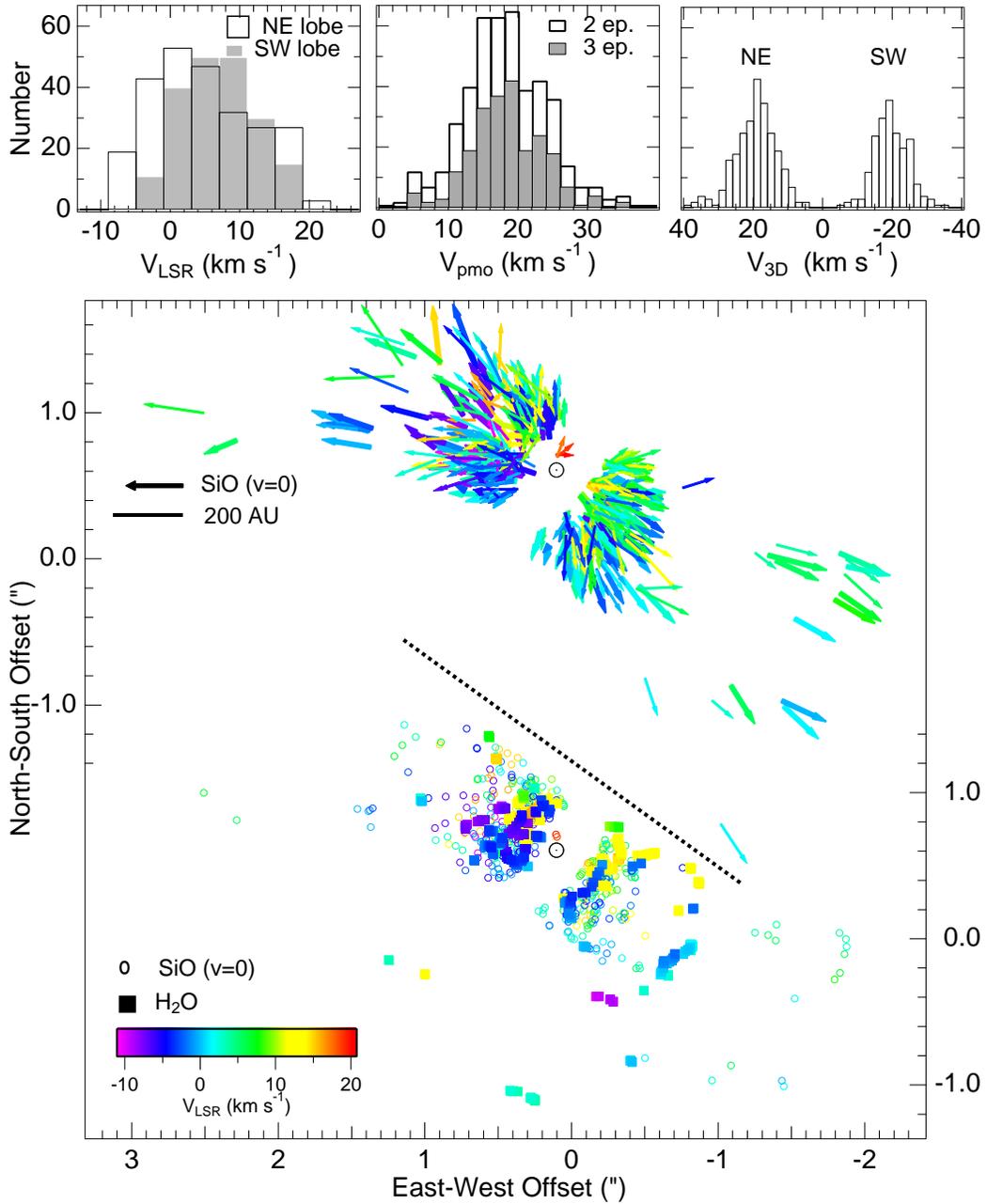}
\caption{Tracers of outflow at radii of 10--1000\,AU from Source I.  
({\it bottom}) Overlay of SiO  $v$=0 ({\it circles}) and H$_2$O 
masers ({\it squares}); the latter appear more concentrated toward 
Source~I ($\odot$)   
({\it middle})  Proper motions for SiO $v$=0 emission clumps, 1999 to
2009.  Proper motions of SiO v=0 emission centroids tracked over 3 or 4 epochs (heavy arrows) and for 2 epochs (light arrows).  
The horizontal black arrow indicates motion of 30\,\kms.  
On both halves of the lower panel, colors indicate V$_{\rm lsr}$ in \kms\ ({\it color bar}); the systemic velocity is 5\,\kms.
({\it upper left})  Distributions of line-of-sight velocities of SiO masers in the northeast and
southwest lobes; their similarity indicates a flow-axis 
close to the sky-plane.   ({\it upper middle})  Histogram of proper 
motions of  SiO spots measured over at least three epochs ({\it shaded})
or two epochs ({\it unshaded}).  ({\it upper right})  Distributions of total space
velocity for SiO maser spots in the northeast and southwest lobes.}

\label{fig:sio_pmo}
\end{figure*}

{\bf H$_2$O--} We correlated pairs of overlapping basebands, stepped to cover V$_{\rm lsr}= -138$ to 137\kms\, ($\nu_{\rm rest}= 22235.08$\,MHz).  We report here on mapping features in the so-called H$_2$O Shell  \citep{genzel81} associated with Source~I.  Each baseband was 1.56\,MHz and channel spacing was 0.16\kms~after Hanning-smoothing. 3C286 and J0530+135  were observed as absolute-flux and bandpass calibrators, respectively.  
One band within each pair was tuned to include the line emission peak near --4.5\kms.
Ringing affected the strongest emission between --4.02 and --5.18\kms.
We flagged these data and used the emission at
--3.86\kms~(1700\,Jy) to obtain  self-calibration solutions every
10$^s$ that were applied to both 1.56\,MHz bands.   
Scans of J0541--056 every 45$^m$ enabled calibration
of instrumental phase offsets between bands.  We detected emission
from $-10.0$ to 16.4\kms, complete to $\sim$1 Jy in each channel, 
except between 8.5 and 11.3\kms~where the completeness limit was restricted to 2--8\,Jy due to dynamic range.

Absolute astrometry was derived from interleaved scans of water maser emission and J0605-085, calibrated using J0541--0541, all observed in dual-polarization continuum mode with 25\,MHz bandwidth.  The estimated absolute position uncertainty is 2\,mas.

\section{Results}

The most intense $v$=0 SiO maser emission occupies two arcs bracketing Source~I, each at a projected radius of $\sim$100\,AU.  This is just outside the maximum radius at which isolated $v$=1 masers are observed (Figure\,\ref{fig:sio_integ_intensity}).  The arcs subtend about the same opening-angle as the nearly radial arms at smaller radii, along which $v$=1,2 maser features are seen to move systematically outward \citep{matthews10}. The northeast arc also overlies in part a 3.78$\mu$m/4.67$\mu$m color temperature minimum \citep{sitarski13}.

\begin{figure*}
\plotone{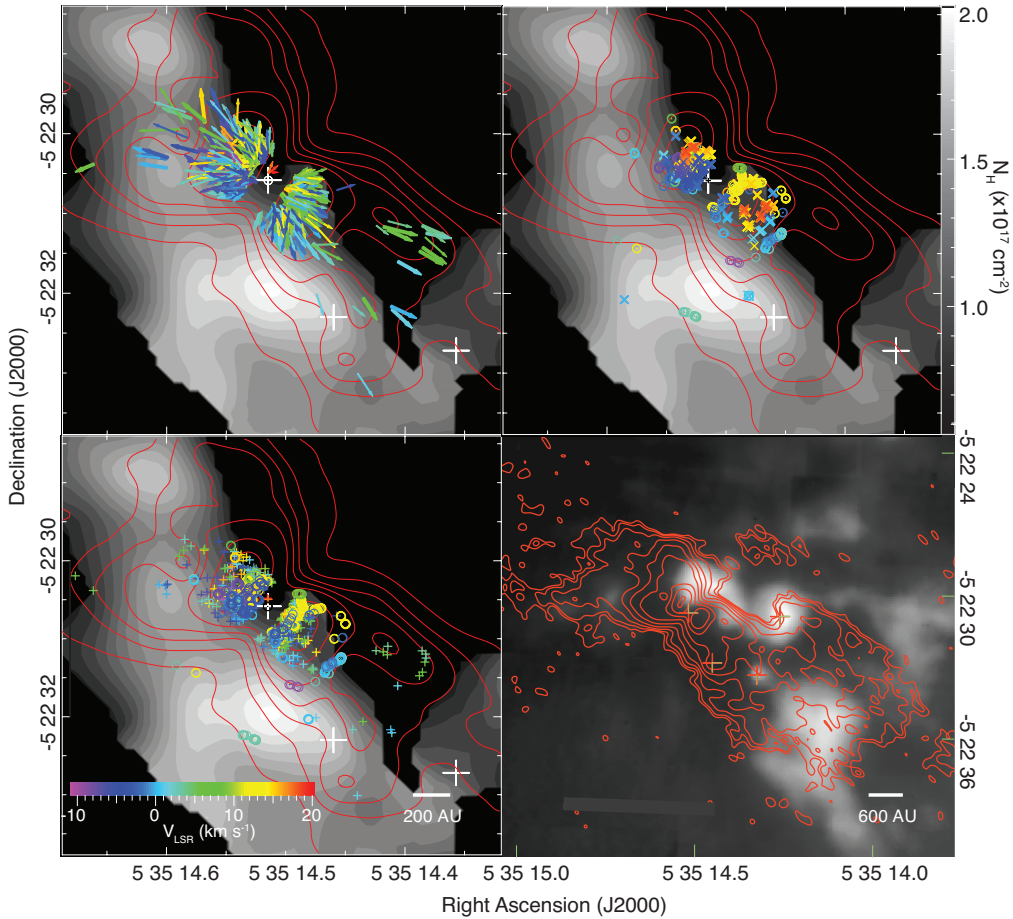}
\vspace{-1.7in}
\caption{
Source I outflow, dense Hot-Core gas, and proximate mid-infrared sources. 
The outflow is shown in velocity-integrated emission of $\lambda 3$\,mm
$J=2\rightarrow1$ emission covering --5 to 5\,\kms, observed with a beamwidth of 0\pas5 ({\it red contours}; 
 \citealt{plambeck2009}). Contours are 3.5, 1.9, 0.55, 0.30, 0.16, 0.086, 0.046, 0.025 Jy bm$^{-1}$ averaged over 10\,\kms.  Only the lower right panel includes the lowest two levels.  Dense gas is shown via an NH$_3$ column-density map with $0\rlap{.}''8$ beamwidth ({\it greyscale}; \citealt{goddi11b}).   The lower-right panel substitutes a mid-infrared image obtained  with Keck (\citealt{g04}). Crosses mark the positions of three YSOs: sources\,I, SMA1, and {\it n} (left to right, as in  \citealt{goddi11b}).  ({\it Upper-left}) Proper motions of outflowing SiO $v$=0 masers ({\it  arrows}). ({\it Upper-right}) Superposition of H$_2$O masers observed in 1983 ($\times$) - \citet{g98}, and 2001 ($\circ$) - this work, demonstrating expansion of the locus. ({\it Lower-left})  Superposition of SiO $v$=0 ($+$) and H$_2$O  masers in 2001 ($\circ$). ({\it Lower-right}) Overlay of the SiO outflow with mid-infrared sources over a larger field-of-view.  
}
\label{fig:sio.pmo.3mm}
\end{figure*}

The angular structure of the $v$=0 emission is suggestive of outflow
in the sky-plane, and its velocity structure confirms it 
(Figure\,\ref{fig:sio_pmo}).  
We tracked proper motions of 59 maser spots for four epochs, 
169 for three epochs, and 219 for two epochs (457 total).  The median proper
motion for maser spots tracked for at least 3 epochs was 18\,\kms.
The corresponding range of 3D velocities in the local frame (V$_{\rm
   LSR}$=5\,\kms) was 4--36\,\kms.  Overlap in the ranges of
radial velocity for the two lobes suggests a close to edge-on geometry (Figure\,\ref{fig:sio_pmo}, upper-left panel).
 Interpretation as an outflow is strengthened by 
 H$_2$O maser emission overlying each lobe of $v$=0 SiO emission
(Figure\,\ref{fig:sio_pmo}). The H$_2$O emission displays a similar
range of line-of-sight velocity (--10.0 to 16.4\,\kms). 
A 20\,\kms~expansion in the angular extent of the H$_2$O distribution
over $\sim$8\,years  \citep{g98} and $\sim$18\,years (Figure\,\ref{fig:sio.pmo.3mm}, lower-left panel) is consistent with the median SiO maser proper motion.

The flow orientation can be estimated from the emission locus  as well as the sky position and proper motions of maser spots.  
We obtain a common mean position angle (PA) of $56^\circ$ by reflecting the southwest lobe about a northwest-southeast line at PA $142^\circ$, which minimizes the standard deviation of the overlapping distributions ($29^\circ$).  
Using the most reliable proper motions (derived from $\ge$3 epochs), the mean motion lies
at a PA$=55\pm34^\circ$ (northeast) and
$-128\pm43^\circ$ (southwest).  
Reflecting the southwest lobe, we obtain a mean outflow PA$=55^\circ$ and a minimum standard 
deviation of $34^\circ$ for a reflection axis of $142^\circ$.  
Hence, we take $56\pm1^\circ$ as the PA of the outflow, measured independently from emission locus and proper motions. 

Although the flow
inside $\sim$100\,AU appears to follow a fixed opening-angle, the
outflow further downstream appears to become more narrowly collimated.
Indeed, the inner quartile range of maser motion position angles at projected radii
$0\rlap{.}''1$--$0\rlap{.}''3$ from Source~I is $80^\circ$, broader than the range of 
$47^\circ$ beyond $0\rlap{.}''3$ (120\,AU).

From our measurements, 
we  estimate the outflow mass-loss rate 
 $\dot{M} = 5 \times 10^{-6} \, V_{18} \, R_{200}^{2} \, n_{6}\Omega/4\pi$ \mbox{\ms~yr$^{-1}$}, where  $V_{18}$  is the average maser velocity in units of 18 \kms, $R_{200}$  is the average distance of SiO masers in units of 200 AU, $n_{6}$ is the volume density in units of  10$^{6}$ cm$^{-3}$, and $\Omega$ is the solid angle for a conical flow. 
The main uncertainty in the formula above is the  density required for excitation  of ground-state SiO masers, known within an order of magnitude  ($10^{6\pm1}$ cm$^{-3}$;  \citealt{goddi09}). 

\begin{figure}
\epsscale{0.91}
\plotone{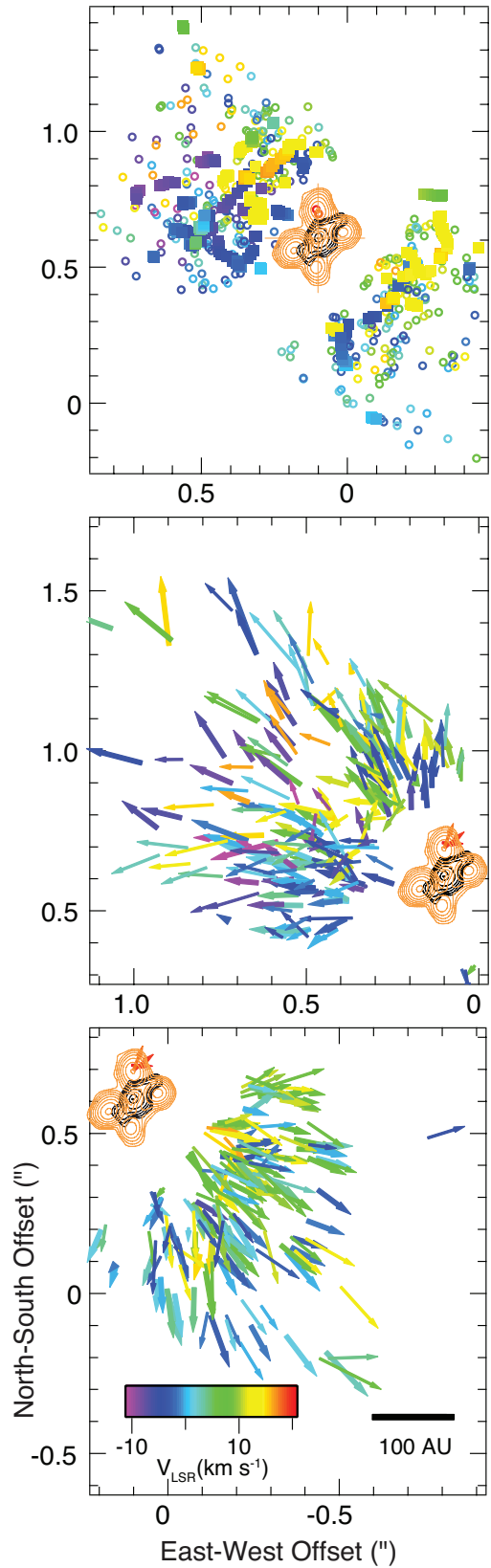}
\caption{Expanded view of the outflow. ({\it top}) SiO $v$=0 and H$_2$O masers ({\it open circles and squares},
   respectively), velocity-integrated SiO $v$=1 masers ({\it red 
contours}), and Source~I $\lambda$7\,mm continuum ({\it black contours}), as mapped with the VLA.
({\it middle}) Expanded view of SiO $v$=0
   maser proper motions in the northeast lobe of Source~I. ({\it
     bottom}) Expanded view of the southwest lobe.}
\label{fig:f2.zoom}
\end{figure}

There is no indication of acceleration/deceleration with radius in the flow. 
But interestingly, in each lobe there is a discernible velocity offset across the minor axis, manifested in the line-of-sight velocities of both SiO and H$_2$O masers (e.g.,  Figure\,\ref{fig:f2.zoom}, upper panel).  Toward the southeast-facing edge, there is a greater preponderance of blueshifted emission; redshifted emission lies preferentially toward the northwest.  The velocity data exhibit a non-Gaussian scatter, so to quantify the trend, we estimate the trimean LSR velocity (the weighted average of median and quartiles) as a function of distance along the minor axis for emission  $0\rlap{.}''1$-$0\rlap{.}''4$ from Source~I:  a 5\kms\ shift for v=0 SiO and a 10\kms\ shift for H$_2$O maser emission.   
We interpret this velocity offset as a signature of rotation parallel to the minor axis of the flow. 

Ground state SiO  $J=1-0$ maser emission  and proper motions displayed in Figures~\ref{fig:sio_integ_intensity}-\ref{fig:f2.zoom} trace only the inner portions of the bipolar outflow traced by  $J=2-1$ emission and mapped with CARMA at $0\rlap{.}''5$ resolution  \citep{plambeck2009}, or the $J=5-4$ emission mapped with ALMA at $1\rlap{.}''5$ resolution \citep{zapata2012,niederhofer2012}.   These transitions show basically the same ``butterfly'' morphology at projected radii $\la$500\,AU and  excellent agreement in the outflow PA ($56^\circ$).   While  complex brightness and velocity-field morphologies are evident well away from Source~I, this may be a consequence of external heating \citep{niederhofer2012}, e.g., by the Hot-Core and compact mid-infrared sources (Figure\,\ref{fig:sio.pmo.3mm}).


\section{Discussion}

\subsection{Outflow 100--1000 AU from Source~I}
The X-shaped morphology traced by vibrationally-excited SiO maser emission  within 100\,AU of Source~I is interpreted as the edges of a bipolar outflow orthogonal to an edge-on rotating disk  \citep{matthews10}.  
Our new mapping of the ground-state SiO and H$_2$O maser emission confirms and extends to 1000\,AU the disk-outflow model. Three lines of evidence support this scenario: (i) the most intense $v$=0 SiO maser emission occupies two arcs that bracket Source~I at a radius of $\sim$100\,AU and subtend an angle corresponding to the opening-angle  of the  vibrationally-excited SiO masers; (ii) the PA of the outflow at radii $>$100 AU is the same as that of the disk and flow axes at small radii; (iii) the line-of-sight velocities of $v=0$ SiO masers indicate an outflow close to the sky-plane, consistent with the nearly edge-on disk.  

Three striking features in the outflow are evident from our measurements: colinearity, recollimation, and rotation. 

Colinearity of northeast and southwest flows ($\S3$) is notable because (i) Source~I lies at the edge of the dense gas associated with the Orion Hot-Core (Figure\,\ref{fig:sio.pmo.3mm}; \citealt{goddi11b}), (ii) the  outflow motion is comparable to the stellar motion, (iii) SiO maser dynamics inside 1000\,AU are indicative of a $\sim$500\,yr crossing time for the outflow, and (iv) the crossing time for Source~I from the center of dynamical interaction with BN  is  also $\sim$500\,yr \citep{goddi11a}, indicating that the onset of flow is contemporaneous with the interaction with BN.  For a hypothetical hydrodynamic flow, the absence of curvature as a result of the YSO motion requires the momentum flux to exceed that of the ambient medium into which Source~I is moving.  Since ground-state SiO maser emission requires densities of 10$^6$-10$^7$\,cm$^{-3}$ (e.g., \citealt{goddi09}), the density of ambient material would have to be $\ll 10^6$\,cm$^{-3}$. However,  ambient gas densities in the vicinity of the Hot-Core are at least this (e.g., \citealt{goddi11b}), thus requiring greater flow energy density than from hydrodynamics alone. 

Outflow (re)collimation is indicated by  a narrower distribution of SiO proper motion position angles far from the YSO, as well as greater preponderance of line-of-sight velocities close to systemic ($\S3$; Figures\,\ref{fig:sio_pmo}, \ref{fig:f2.zoom}).   Mechanical collimation from the ambient medium is rendered problematic by the similarity of the leading and trailing edges of the flow despite the anticipated ambient density gradient toward the Hot-Core.  In principle, maser excitation effects could bias the inferred morphology of the flow if these favor emission close to the outflow axis, but this would not explain comparable leading and trailing edge gradients in the intensity of thermal SiO emission (e.g., Figure\,\ref{fig:sio.pmo.3mm}, top).

Finally, there is a discernible rotation signature about the major axis of the flow in each lobe, consistent with the rotation observed at radii of tens of AU in the vibrationally-excited SiO masers \citep{matthews10}.
Our data  are suggestive of these dynamics being communicated from scales of O(10)\,AU to at least O(100)\,AU.

\subsection{Magnetocentrifugal Wind from a high-mass YSO}
\label{magwind}
The possibility of a magnetohydrodynamic disk-wind is raised by the evidence of a rotating wide-angle outflow launched from a compact disk, that is recollimated downstream, and that proceeds undeflected through a dense medium.  Rotation is anticipated for a magnetized outflow with the field anchored to a rotating disk.  Magnetic field lines threading the flow would  raise its energy density, while a  toroidal field and corresponding hoop stress generated by rotation could efficiently narrow collimation with distance.

\citet{matthews10} conservatively interpreted the maser data in the context of Keplerian motion and the dominant action of gravity, inferring a dynamical mass of $\sim$8\,M$_{\odot}$.  However,  early indirect evidence of  non-gravitational effects were noted (e.g., curved maser trajectories), possibly due to magnetic fields, by which rotation would appear Keplerian though the YSO mass  would be underestimated.  The latter is consistent with the difference between YSO masses inferred by \citet{matthews10} and  \citet{goddi11a} under the assumption that BN and Source~I are in recoil \citep[cf.][]{tan12}.  
This early evidence, along with the morphology and dynamics of outflow on scales out to 1000\,AU, strengthens the case for a magnetic flow.

Using axisymmetric MHD numerical simulations, \citet{vaidya13} have explored the plausibility of an MHD origin of the wide-angle flow probed by vibrationally-excited SiO masers inside 100\,AU from Source~I,  and proposed that the SiO masers may be excited as an MHD driven wind interacts with the ambient molecular medium in form of shocks.
\citet{seifried12} studied earlier conditions, applying MHD theory from the collapse of magnetized cloud cores to disk formation and outflow launching, and demonstrated magnetocentrifugal launching of massive outflows, similar to the case of low-mass outflows. 

Why do intense ground-state SiO and H$_2$O maser emission arise suddenly at 100 AU?  Why are SiO and H$_2$O masers apparently intermixed when the densities required for emission differ by (conservatively) an order of magnitude? For a YSO luminosity of 2$\times$10$^4$ L$_\odot$ (i.e., a binary with two 10\,M$_\odot$ stars), the sublimation radius is $\ll$100\,AU, and since maser emission requires a high gas-phase abundance, its appearance  so far out in the flow is significant.

We propose that the arcs of maser emission at $\sim$100\,AU radius indicate  the onset of strong shocks in dusty outflowing material.  Hydromagnetic C-type shocks as slow as 10-20\,km\,s$^{-1}$ (comparable to the flow speed) are capable of sputtering grains \citep[][and references therein]{schilke1997,loo13}, a process that would raise the gas-phase abundance of SiO and H$_2$O.  Formation of two continuous shock structures subtending broad ranges of polar angle and narrow ranges in  radius indicates a systematic change in physical conditions.   Transition to a super-Alfv\'enic flow and consequent shock formation may trigger the observed (re)appearance of maser emission in the outflow at $\sim$100\,AU.  
Decline in Alfv\'en velocity below the outflow velocity would require the magnetic field to decline at least linearly with radius if density falls quadratically.  This is not implausible.   An observational consequence is that the inner edge of the maser emission locus would not appear to expand with time.   

The presence of maser emission well downstream, suggests persistent high gas-phase abundance, as well as energy that can drive maser pumping.   Flow speeds in excess of the sound and Alfv\'en speeds would drive shocks and impart pump energy over a wide range of radii, assuming that cooling timescales are much shorter than dynamical time scales.  In this region, observed fading of the velocity gradient indicative of rotation around the flow axis is consistent with decoupling of the neutral gas from the field as expected from MHD disk-wind models.

\section{Conclusions}

Data for SiO and H$_2$O masers  provide an unusually detailed view of the  launch and collimation of an outflow from the surface of a compact  disk surrounding Source~I in BN/KL.   Position and velocity resolved gas dynamics at projected radii of 10\,AU to 1000\,AU  suggest the presence of a magnetocentrifugal disk-wind driven by a massive YSO. This is notable in view of continuing ambiguity concerning the role of magnetic fields in high-mass star formation.  
While the outflow structure up to 1000 AU in this high-mass YSO is remarkable in terms of symmetry, colinearity, collimation, and rotation, larger scales reveal the effects of the interaction of the massive outflow  with the typically complex environment of a massive star forming region.


\begin{thebibliography}{}

\bibitem[Blandford \& Payne (1982)] {bp82} Blandford, R. D. \& Payne, D. G. 1982, MNRAS, 199, 883

\bibitem[Chatterjee 
\& Tan(2012)]{tan12} Chatterjee, S., \& Tan, J.~C.\ 2012, \apj, 754, 152 

\bibitem[Genzel et al.(1981)]{genzel81} Genzel, R., Reid, M.~J.,
Moran, J.~M., \& Downes, D.\ 1981, \apj, 244, 884

\bibitem[Gezari et al.(1998)]{gezari98} Gezari, D. Y., Backman, D. E., 
\& Werner, M. W.\
1998, ApJ, 509, 283

\bibitem[Goddi et al.(2009)]{goddi09} Goddi, C., Greenhill,
L.~J., Chandler, C.~J., Humphreys, E.~M.~L., Matthews, L.~D.,
\& Gray, M.~D.\ 2009, \apj, 698, 1165

\bibitem[Goddi et al.(2011a)]{goddi11a} Goddi, C., Humphreys, E.~M.~L., 
Greenhill,
L.~J., Chandler, C.~J., Matthews, L.~D.\ 2011a, \apj, 728, 15

\bibitem[Goddi et al.(2011b)]{goddi11b} Goddi, C., Greenhill, 
L.~J., Humphreys, E.~M.~L., Chandler, C.~J., 
\& Matthews, L.~D.\ 2011b, \apjl, 739, L13 

\bibitem[Greenhill\etal (1998)]{g98} Greenhill, L. J., Gwinn, C. R., 
Schwartz, C.,
Moran, J. M., \& Diamond, P. J. 1998, Nature, 396, 650

\bibitem[Greenhill et al.(2004a)]{g04} Greenhill, L. J.,  Gezari, D. Y., 
Danchi, W. C., Najita, J., Monnier, J. D., \& Tuthill,  P. G.\ 2004a, ApJ, 605, L57

\bibitem[Greenhill et al.(2004b)]{Gre04b} Greenhill, L.~J., 
Reid, M.~J., Chandler, C.~J., Diamond, P.~J., 
\& Elitzur, M.\ 2004b, Star Formation at High Angular Resolution, 221, 155

\bibitem[Kim et al.(2008)]{kim08} Kim, M.~K., et al.\ 2008,
\pasj, 60, 991

\bibitem[K\"onigl \& Pudritz (2000)]{kp00} K\"onigl, A. \& Pudritz, R. E. 2000, Protostars and Planets IV, ed. V. Mannings, A. P. Boss, \& S. S. Russell (Tucson, AZ: Univ. Arizona Press), 759

\bibitem[Matthews et al.(2010)]{matthews10} Matthews, L.~D.,
Greenhill, L.~J., Goddi, C.,  Chandler, C.~J., Humphreys, E.~M.~L., \& 
Kunz, M.  2010, \apj, 708, 80

\bibitem[Niederhofer et 
al.(2012)]{niederhofer2012} Niederhofer, F., Humphreys, E.~M.~L., \& Goddi, C.\ 2012, \aap, 548, A69 

\bibitem[Plambeck et al.(2009)]{plambeck2009} Plambeck, R.~L., 
Wright, M.~C.~H., Friedel, D.~N., et al.\ 2009, \apjl, 704, L25 

\bibitem[Plambeck et al.(2013)]{plambeck2013} Plambeck, R.~L., 
Bolatto, A.~D., Carpenter, J.~M., et al.\ 2013, \apj, 765, 40 

\bibitem[Reid et al.(2007)]{reid07} Reid, M.~J., Menten, K.~M., Greenhill, L.~J., \& Chandler, C.~J.\ 2007, \apj, 664, 950

\bibitem[Schilke et 
al.(1997)]{schilke1997} Schilke, P., Walmsley, C.~M., Pineau des Forets, G., \& Flower, D.~R.\ 1997a, \aap, 321, 293 

\bibitem[Seifried et al.(2012)]{seifried12} Seifried, D., Pudritz, 
R.~E., Banerjee, R., Duffin, D., \& Klessen, R.~S.\ 2012, \mnras, 422, 347 

\bibitem[Shuping et al.(2004)]{shuping04} Shuping, R. Y., Morris, M., \& 
Bally, J.\ 2004,
AJ, 128, 363

\bibitem[Sitarski et al.(2013)]{sitarski13} Sitarski, B. N., 
Morris, M. R., Lu, J. R., Duch\^{e}ne, G., Stolte, A., Becklin, E. E., Ghez, A., Zinnecker, H.  2013, \apj, in press

\bibitem[Vaidya 
\& Goddi(2013)]{vaidya13} Vaidya, B., \& Goddi, C.\ 2013, \mnras, 429, L50 

\bibitem[Van Loo et al.(2013)]{loo13} Van Loo, S., Ashmore, 
I., Caselli, P., Falle, S.~A.~E.~G., 
\& Hartquist, T.~W.\ 2013, \mnras, 428, 381 

\bibitem[Zapata et al.(2012)]{zapata2012} Zapata, L.~A., 
Rodr{\'{\i}}guez, L.~F., Schmid-Burgk, J., et al.\ 2012, \apjl, 754, L17

\end{thebibliography}
\end{document}